\documentclass[twocolumn,aps,prd,amsmath,amssymb]{revtex4}

\usepackage{bm}
\usepackage{amsmath}
\usepackage{amssymb}
\usepackage{latexsym}
\usepackage{amsfonts}
\usepackage{epsfig}
\usepackage{psfrag}
\usepackage{graphicx}

\usepackage{latexsym}
\usepackage{graphics}
\usepackage{amsmath}
\usepackage{amssymb}
\usepackage{latexsym}
\usepackage{amsfonts}
\usepackage{epsfig}
\usepackage{psfrag}
\usepackage{graphicx}
\usepackage{slashed}

\newcommand{\ket}[1]{\left|#1\right>}

\newcommand{\nn}{\nonumber\\}

\newcommand{\f}[1]{\mbox{\boldmath$#1$}}

\newcommand{\bea}{\begin{eqnarray}}
\newcommand{\ea}{\end{eqnarray}}
\newcommand{\eea}{\end{eqnarray}}
\newcommand{\ord}{\,{\cal O}}
\newcommand{\asl}{\slashed{a}}
\newcommand{\psl}{\slashed{p}}
\newcommand{\ksl}{\slashed{k}}

\begin{document}

\title{Quantum radiation by electrons in lasers and the Unruh effect}

\author{Ralf Sch\"utzhold and Clovis Maia}

\affiliation{Fachbereich Physik, Universit\"at Duisburg-Essen, 
D-47048 Duisburg, Germany}
%Insert the first address here \and the second here}
%
%\date{Received: date / Revised version: date}
% The correct dates will be entered by Springer
%
\abstract{
In addition to the Larmor radiation known from classical electrodynamics, 
electrons in a laser field may emit pairs of entangled photons -- which 
is a pure quantum effect. 
We investigate this quantum effect and discuss why it is suppressed 
in comparison with the classical Larmor radiation (which is just 
Thomson backscattering of the laser photons). 
Further, we provide an intuitive  explanation of this process 
(in a simplified setting) in terms of the Unruh effect.  
%
%Insert your abstract here.
%
%\PACS{
%{04.62.+v}{Quantum field theory in curved spacetime}   \and
%{12.20.Fv}{Experimental tests of QED}    \and
%{41.60.-m}{Radiation by moving charges}    \and
%{42.50.Dv}{Nonclassical states of the electromagnetic field}
%, including
	 %entangled photon states; quantum state engineering and measurements
	 %(see also 03.65.Ud Entanglement and quantum nonlocality,
	 %e.g. EPR paradox, Bell's inequalities, GHZ states, etc.)
%      {PACS-key}{discribing text of that key}   \and
%      {PACS-key}{discribing text of that key}
     } % end of PACS codes
} %end of abstract
\maketitle
%

%%%%%%%%%%%%%%%%%%%%%%%%%%%%%%%%%%%%%%%%%%%%%%%%%%%%%%%%%%%%%%%%%%%%%%%%%%%%%%%
\section{Introduction}\label{intro}
%%%%%%%%%%%%%%%%%%%%%%%%%%%%%%%%%%%%%%%%%%%%%%%%%%%%%%%%%%%%%%%%%%%%%%%%%%%%%%%

%November 17, 2008; max. 20 pages 

\begin{figure}
\centerline{\resizebox{0.25\textwidth}{!}{\includegraphics{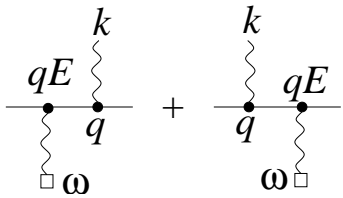}}}
\caption{Lowest-order Feynman diagrams for the classical Larmor radiation,
which is just Thomson (or Compton) scattering of the incoming laser photons.
The horizontal straight line denotes the electron and the vertical wiggly 
lines are the photons. The external laser field of frequency $\omega$ is 
indicated by a square and its vertex (black dot) scales with $qE$. 
The emitted photon has momentum $k$ and its vertex just scales with $q$.}
\label{feynman1}
\end{figure}

With the availability of strong and stable lasers in a very clean background, 
we are rapidly approaching a regime in which relativistic quantum effects start 
to play a role.
As a simple example, let us consider an electron (of mass $m$ and charge $q$) 
in a laser field (which we approximate by a plane wave). 
Here we assume that the laser has a small (e.g., optical) frequency $\omega$ 
and a large (but not too large) Keldysh adiabaticity parameter 
(i.e., small electric field strength $E$ when normalized w.r.t.~$m$)
\bea
\label{kel}
\omega\ll m
\,,
\quad
\gamma=\frac{m\omega}{qE}\gg1
\,. 
\ea
In this case, the electron oscillates non-relativistically and nearly 
harmonically and thus emits radiation of the laser frequency -- 
which is just the Larmor radiation known from classical electrodynamics. 
Alternatively, one can view this classical process as Thomson
(i.e., low-energy Compton) backscattering of the laser photons.
The lowest-order Feynman diagrams of this process are depicted 
in Fig.~\ref{feynman1}.
Due to $\omega\ll m$, the recoil of the electron can carry away momentum, 
but basically no energy $\f{p}^2_{\rm out}/m\ll\omega$ and thus we 
can directly read off the classical resonance condition 
$|\f{k}|=\omega$ (energy conservation). 

%%%%%%%%%%%%%%%%%%%%%%%%%%%%%%%%%%%%%%%%%%%%%%%%%%%%%%%%%%%%%%%%%%%%%%%%%%%%%%%
%%%%%%%%%%%%%%%%%%%%%%%%%%%%%%%%%%%%%%%%%%%%%%%%%%%%%%%%%%%%%%%%%%%%%%%%%%%%%%%

However, in addition to these lowest-order Feynman diagrams in 
Fig.~\ref{feynman1}, there are also further (next-to-leading-order) 
Feynman diagrams such as the one sketched in Fig.~\ref{feynman3}.
They correspond to the emission of two photons with momenta $\f{k}_1$ 
and $\f{k}_2$.
With the same argument as before, the resonance condition 
(energy conservation) now reads $|\f{k}_1|+|\f{k}_2|=\omega$, 
i.e., the sum of the energies of the two emitted photons equals 
the laser frequency.  
However, if the electron oscillates with frequency $\omega$, the 
emission of two photons with, say, half the energy each 
$|\f{k}_1|=|\f{k}_2|=\omega/2$ cannot be explained within classical 
electrodynamics.
It is, therefore, a pure quantum effect.
This interesting observation poses two major questions:
\begin{itemize}
 \item How can this effect be understood? 
 \item One would expect that this quantum effect is suppressed 
       compared to the classical process in Fig.~\ref{feynman1}. \\
       If this is correct, how does this suppression work? 
\end{itemize}
In the following, we shall discuss these two questions. 

%%%%%%%%%%%%%%%%%%%%%%%%%%%%%%%%%%%%%%%%%%%%%%%%%%%%%%%%%%%%%%%%%%%%%%%%%%%%%%%
\section{Classical Radiation}\label{class}
%%%%%%%%%%%%%%%%%%%%%%%%%%%%%%%%%%%%%%%%%%%%%%%%%%%%%%%%%%%%%%%%%%%%%%%%%%%%%%%

Let us first discuss the classical process, which is just Larmor radiation 
known from classical electrodynamics and can be understood as Thomson 
(or Compton) scattering of the incoming laser photons by the electron. 
We employ standard notation with $a\cdot p=a_\mu p^\mu$ and 
$\asl=a_\mu\gamma^\mu$ etc. 
Here $p_{\rm in}$ and $p_{\rm out}$ are the initial and final four-momenta 
of the electron, respectively, and $u_{\rm in}$ as well as $u_{\rm out}$
are the associated spinor solutions.
The amplitude and momentum of the laser photon are denoted by 
$a_L$ and $k_L$, whereas $a_1$ and $k_1$ describe the emitted photon. 
The first Feynman diagram in Fig.~\ref{feynman1} generates the amplitude 
\bea
{\cal A}_{\rm class}=
\bar u_{\rm in}\,q\asl_L\,\frac{1}{\psl_{\rm in}+\ksl_L-m}\,q\asl_1\,u_{\rm out}
\,.
\ea
Since $p_{\rm in}$ and $k_L$ are on-shell, the square of the denominator in the 
electron propagator yields $2p_{\rm in}\cdot k_L$. 
In the following, we go into the (initial) rest frame of the electron 
$p_{\rm in}=(m,0)$ which yields $2p_{\rm in}\cdot k_L=2m\omega_L$.
Furthermore, we perform an expansion into (inverse) powers of the electron 
mass $m$ where $\sim$ denotes the scaling in leading order 
\bea
{\cal A}_{\rm class}\sim\frac{q^2}{m\omega_L}\,
\bar u_{\rm in}\,\asl_L(\psl_{\rm in}+\ksl_L+m)\asl_1\,u_{\rm out}
\,.
\ea
Using the Clifford algebra 
$\asl_L\psl_{\rm in}+\psl_{\rm in}\asl_L=2p_{\rm in}\cdot a_L$
and $\bar u_{\rm in}(\psl_{\rm in}-m)=0$, we get 
\bea
{\cal A}_{\rm class}\sim\frac{q^2}{m\omega_L}\,
\bar u_{\rm in}
(\asl_L\ksl_L\asl_1+2p_{\rm in}\cdot a_L\asl_1)
u_{\rm out}
\,.
\ea
In leading order in $1/m$, we may neglect the recoil of the electron 
$p_{\rm out}\approx p_{\rm in}$ and set $u_{\rm out}\approx u_{\rm in}$ 
which leads to 
\bea
{\cal A}_{\rm class}\sim\frac{q^2}{m\omega_L}\,
{\rm Tr}\left\{
(\asl_L\ksl_L\asl_1+2p_{\rm in}\cdot a_L\asl_1)(\psl_{\rm in}+m)
\right\}
\,.
\ea
In temporal (radiation) gauge, we have $p_{\rm in}\cdot a_L=0$ 
as well as $p_{\rm in}\cdot a_1=0$, which yields the final result 
\bea
{\cal A}_{\rm class}\sim q^2 a_1\cdot a_L
+\ord\left(\frac{1}{m}\right)
\,.
\ea
This expression reproduces the well-known result that the emitted photon 
has the same polarization as the laser field, i.e., in the direction of
the electron trajectory. 
(One obtains the same result for the other lowest-order diagram in 
Fig.~\ref{feynman1}.) 

%%%%%%%%%%%%%%%%%%%%%%%%%%%%%%%%%%%%%%%%%%%%%%%%%%%%%%%%%%%%%%%%%%%%%%%%%%%%%%%
\section{Quantum Radiation}\label{quant}
%%%%%%%%%%%%%%%%%%%%%%%%%%%%%%%%%%%%%%%%%%%%%%%%%%%%%%%%%%%%%%%%%%%%%%%%%%%%%%%

%%%%%%%%%%%%%%%%%%%%%%%%%%%%%%%%%%%%%%%%%%%%%%%%%%%%%%%%%%%%%%%%%%%%%%%%%%%%%%%
\begin{figure}
\centerline{\resizebox{0.175\textwidth}{!}{\includegraphics{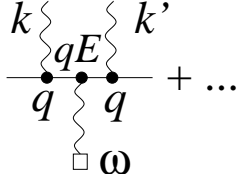}}}
\caption{One of the (six) lowest-order Feynman diagrams for quantum radiation. 
The notation is the same as in Fig.~\ref{feynman1}.}
\label{feynman3}
\end{figure}
%%%%%%%%%%%%%%%%%%%%%%%%%%%%%%%%%%%%%%%%%%%%%%%%%%%%%%%%%%%%%%%%%%%%%%%%%%%%%%%

Now, having discussed the classical process, let us turn to the 
quantum radiation, cf.~Fig.~\ref{feynman3}.
In this case, we have two emitted photons with momenta $k_1$ and $k_2$ 
as well as amplitudes $a_1$ and $a_2$. 
The Feynman diagram in Fig.~\ref{feynman3} then generates the amplitude 
\bea
{\cal A}_{\rm quant}=q^3
\bar u_{\rm in}\,\asl_1\,
\frac{1}{\psl_{\rm in}-\ksl_1-m}
\,\asl_L\,
\frac{1}{\psl_{\rm out}-\ksl_2-m}\,
\asl_2\,u_{\rm out}
\,. 
\nn
\ea
With manipulations analogous to those in the previous Section, 
we find the leading-order behavior 
\bea
{\cal A}_{\rm quant}\sim\frac{q^3}{m^2\omega_1\omega_2}\,
\bar u_{\rm in}
(2p_{\rm in}\cdot a_1-\asl_1\ksl_1)
\asl_L
\times
\nn
\times
(2p_{\rm out}\cdot a_2-\asl_2\ksl_2)
u_{\rm out}
\,.
\ea
Remembering $p_{\rm in}\cdot a_1=\ord(m^0)$ and
$p_{\rm out}\cdot a_2=\ord(m^0)$, we see that the amplitude 
for quantum radiation is suppressed by a factor of $1/m$.
Again neglecting the recoil of the electron and setting 
$u_{\rm out}\approx u_{\rm in}$, we get 
\bea
{\cal A}_{\rm quant}\sim\frac{q^3}{m^2\omega_1\omega_2}\,
{\rm Tr}\left\{
%(\asl_1\ksl_1+2p_{\rm in}\cdot a_1)
(2p_{\rm in}\cdot a_1-\asl_1\ksl_1)
\asl_L
\right.
\times
\nn
\times
\left.
%(\asl_2\ksl_2+2p_{\rm out}\cdot a_2)
(2p_{\rm out}\cdot a_2-\asl_2\ksl_2)
(\psl_{\rm in}+m)
\right\}
\,.
\ea
The trace of the products of $\gamma^\mu$-matrices yields various 
combinations of scalar products of the four-vectors involved.
The leading terms (in $1/m$) are obtained when $\psl_{\rm in}$
is combined with $\ksl_1$ or $\ksl_2$ which yields  
$p_{\rm in}\cdot k_{1,2}=m\omega_{1,2}+\ord(m^0)$.
Considering the case where $\psl_{\rm in}$ is combined with 
$\ksl_1$, the remaining contributions are scalar products 
of the four-vectors $a_1$, $a_2$, $a_L$, and $k_2$.
Thus we obtain a polarization entangled part 
\bea
{\cal A}_{\rm quant}^{\rm entangl}\sim\frac{q^3}{m}\,
(a_1\cdot a_2)(k_1\cdot a_L)+
1\leftrightarrow2
%\left(
%\frac{k_1\cdot a_L}{\omega_1}
%+\frac{k_2\cdot a_L}{\omega_2}
%\right)
%???
\,,
\ea
plus a remaining part with non-entangled polarizations, 
which  contains contributions such as 
%$(a_1\cdot a_L)(a_2\cdot k_1)$ and 
$(a_1\cdot a_L)(a_2\cdot p_{\rm out})$ as well as 
the interchanged terms $1\leftrightarrow2$. 
%
%(Remember that $a_1\cdot k_1=a_2\cdot k_2=0$.)

In the polarization entangled part ${\cal A}_{\rm quant}^{\rm entangl}$,
the polarizations of the two emitted photons are equal, whereas the 
polarization of each one is undetermined, i.e., we obtain an EPR-like 
state in the polarization
\bea
\label{EPR}
\ket{{\rm EPR}}=
\frac{\ket{\updownarrow\updownarrow}+\ket{\leftrightarrow\leftrightarrow}}
{\sqrt{2}}
\ea
In the remaining contributions, the polarizations of the two outgoing
photons are determined separately by the laser field and the momenta 
involved.
If both photons propagate in the same direction as the electron 
(given by laser polarization $a_L$), these remaining terms vanish and 
only the entangled part ${\cal A}_{\rm quant}^{\rm entangl}$ survives. 

%%%%%%%%%%%%%%%%%%%%%%%%%%%%%%%%%%%%%%%%%%%%%%%%%%%%%%%%%%%%%%%%%%%%%%%%%%%%%%%
\section{Unruh Effect}\label{unruh}
%%%%%%%%%%%%%%%%%%%%%%%%%%%%%%%%%%%%%%%%%%%%%%%%%%%%%%%%%%%%%%%%%%%%%%%%%%%%%%%

%%%%%%%%%%%%%%%%%%%%%%%%%%%%%%%%%%%%%%%%%%%%%%%%%%%%%%%%%%%%%%%%%%%%%%%%%%%%%%%
\begin{figure}
\centerline{\resizebox{0.25\textwidth}{!}{\includegraphics{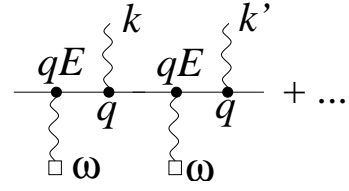}}}
\caption{One of the lowest-order Feynman diagrams for two-photon 
Thomson (Compton) scattering, which is a purely classical effect. 
Again, the notation is the same as in Fig.~\ref{feynman1}.}
\label{feynman2}
\end{figure}
%%%%%%%%%%%%%%%%%%%%%%%%%%%%%%%%%%%%%%%%%%%%%%%%%%%%%%%%%%%%%%%%%%%%%%%%%%%%%%%

With the above estimates, we can already answer the second question posed 
in the Introduction, i.e., the suppression of the quantum effect in 
Fig.~\ref{feynman3} compared with the classical Larmor radiation, 
cf.~Fig.~\ref{feynman1}. 
Of course -- as one can simply infer from the Feynman diagrams in 
Figs.~\ref{feynman1} and \ref{feynman3} by counting the vertices -- 
the probability ${\cal P}_{\rm quant}\sim|{\cal A}_{\rm quant}^2|$
of the quantum effect contains one factor of 
$\alpha_{\rm QED}\approx1/137$ (i.e., the fine-structure constant) 
more than the classical 
probability ${\cal P}_{\rm class}\sim|{\cal A}_{\rm class}^2|$.
However, the power of $\alpha_{\rm QED}$ is not sufficient for 
distinguishing classical and quantum effects:
As a counter-example, consider the process of two-photon scattering 
sketched in Fig.~\ref{feynman2}, which scales with $\alpha_{\rm QED}^4$
but is still a purely classical effect.
Within our approach, the expected suppression of the quantum effect 
manifests itself in the small ratio $\omega/m$, i.e., 
\bea
\frac{{\cal P}_{\rm quant}}{{\cal P}_{\rm class}}
=
\ord\left(\alpha_{\rm QED}\left[\frac{\hbar\omega}{mc^2}\right]^2\right)
\ll1
\,.
\ea
Here, we inserted $\hbar$ and $c$ in order to illustrate that we 
are dealing with a relativistic quantum effect $\sim\hbar/c^2$.
Intuitively speaking, the heavier the electron is, the more 
classically it behaves. 

Having addressed the problem of suppression, let us now turn
to the first question posed in the Introduction: 
How can we understand this effect?
As a first approach, let us simplify the situation by assuming
that the two emitted photons propagate in the same direction as 
the electron, i.e., $\f{k}_1\|\f{k}_2\|\f{e}_L$.
In this case, we have an effectively 1+1-dimensional situation,
where the two polarizations can be treated as two independent 
scalar fields 
$\phi_\leftrightarrow(t,x)$ and $\phi_\updownarrow(t,x)$. 
%
%$\phi_\circlearrowright(t,x)$ and 
%$\phi_\circlearrowleft(t,x)$. 
%
%Here, we have used the circular polarizations $\circlearrowright$ 
%and $\circlearrowleft$ in order to facilitate the discussion 
%of angular momentum conservation.
%
As a further simplification, we replace the harmonic oscillation 
of the electron by a uniform acceleration $\mathfrak a$. 
In this case, an observer sitting on the accelerated electron 
will experience the Minkowski vacuum as a thermal bath with 
the Unruh temperature \cite{Unruh-prd}
\bea
T_{\rm Unruh}=\frac{\hbar\,{\mathfrak a}}{2\pi k_{\rm B}c}
\,.
\ea
In view of the finite cross section $\sigma\propto q^4/m^2$
of the electron for Thomson scattering, this uniformly accelerated 
observer would conclude that there is a finite probability for
scattering a photon out of this thermal bath back into another mode
with the same polarization.  
This scattering event would change the thermal bath experienced 
by the accelerated observer since it removes a photon from one mode 
and adds it to another mode. 
However, since this thermal bath in the accelerated frame is just 
our Minkowski vacuum in the inertial frame, this modification must 
also change the Minkowski vacuum:
The removal of a thermal photon in the accelerated frame corresponds 
to the emission of a real photon in the inertial frame
(since the Minkowski vacuum is the ground state, changing this state 
can only be done via an excitation) and the re-insertion of the 
thermal photon into another mode does also correspond to the emission 
of a real photon in the inertial frame \cite{Happens}. 
As a result, this scattering event in the accelerated frame translates
to the emission of two real photons in the inertial frame
\cite{Habs}. 
The expected scaling behavior from the Thomson cross section 
\bea
\label{scaling}
{\cal P}_{\rm quant}\sim\sigma\times f(T_{\rm Unruh})\sim
\frac{q^4}{m^2}\times f(qE)
\ea
fits this simple picture. 
In this way, we can understand the relativistic quantum effect 
sketched in Fig.~\ref{feynman3} in terms of the Unruh effect -- 
after some very strong simplifications\footnote{Nevertheless, even 
for non-uniform accelerations, the accelerated observer would still 
experience excitations -- though no longer in a stationary thermal 
state -- so the basic picture remains correct.}.

In order to facilitate the discussion of angular momentum 
conservation, it is more convenient to use circular polarizations 
$\circlearrowright$ and $\circlearrowleft$.
In this basis, the two fields $\phi_\circlearrowright(t,x)$ and 
$\phi_\circlearrowleft(t,x)$ are roughly equivalent to the carriers 
of positive and negative charge, respectively, and the EPR state 
in Eq.~(\ref{EPR}) reads 
\bea
\ket{{\rm EPR}}=
\frac{\ket{\circlearrowleft\circlearrowright}-
\ket{\circlearrowright\circlearrowleft}}
{\sqrt{2}}
\,.
\ea
Within this picture, the entanglement of the above Bell state can 
be understood as a consequence of the vacuum entanglement, which 
is responsible for the thermal nature of the Unruh effect:
The uniformly accelerated observer can only access a part of the 
full Minkowski space-time and hence experiences the pure state
of the Minkowski vacuum as a mixed state (thermal density matrix). 

%%%%%%%%%%%%%%%%%%%%%%%%%%%%%%%%%%%%%%%%%%%%%%%%%%%%%%%%%%%%%%%%%%%%%%%%%%%%%%%
\section{Conclusions and Outlook}\label{concl}
%%%%%%%%%%%%%%%%%%%%%%%%%%%%%%%%%%%%%%%%%%%%%%%%%%%%%%%%%%%%%%%%%%%%%%%%%%%%%%%

In summary, we have studied the process described by the Feyman diagrams 
sketched in Fig.~\ref{feynman3} and concluded that it is a pure 
(relativistic) quantum effect, which cannot be explained within 
classical electrodynamics \cite{Thirolf}. 
As one would expect, it is suppressed (in the parameter region under 
consideration) in comparison with the classical Larmor radiation
(cf.~Figs.~\ref{feynman1} and \ref{feynman2}).
This suppression manifests itself by an additional power of $1/m$
(rather than $\alpha_{\rm QED}$, cf.~Fig.~\ref{feynman2}).
As one would expect, heavier particles behave more classically than 
light ones (if all the other parameters remain the same). 

Furthermore, we have provided an intuitive explanation of this quantum 
process in terms of the Unruh effect using a very simplified 1+1 
dimensional geometry, i.e., in forward direction 
$\f{k}_1\|\f{k}_2\|\f{e}_L$.
In other directions, there are also non-entangled contributions 
and the situation becomes more complicated. 
One reason for this lies in the fact that the polarization can no 
longer be treated as a charge which is independent of other 
quantum numbers. 
The question of whether the full 3+1 dimensional quantum effect can 
also be understood in terms of the Unruh effect or is due to a 
different quantum mechanism requires further study. 

Interestingly, there is a paper more than twenty years old 
\cite{Zelddovich} by Zel'dovich {\em et al}, in which they discuss 
the same major idea and conclude that a scattering event in the 
accelerated frame corresponds to the emission of two real photons 
in the laboratory frame 
(even though they did not discuss their entanglement).
In addition, they estimated the pair creation rate in analogy to 
Eq.~(\ref{scaling}). 
However, apparently this paper did not receive as much attention 
as it deserved and in a later paper \cite{Chen+Tajima}, 
Chen and Tajima discussed the signatures of the Unruh effect 
without noting its two-photon 
nature\footnote{We would like to stress that this remark is not 
meant to blame Chen and Tajima for not knowing the paper by 
Zel'dovich {\em et al} -- we also just discovered it recently.}.  

It should also be noted that the Feyman diagrams sketched in 
Fig.~\ref{feynman3} have been studied extensively before in 
connection with double Compton scattering \cite{double}, 
but mainly from a different point of view.  
For example, many of these calculations are based on an avergage 
over the photon polarizations -- whereas here we are specifically 
interested in the entanglement in polarization and its relation to 
the quantum nature of the process\footnote{It might be adequate to 
add a few comments regarding the recent paper 
{\tt arXiv:0809.1505v1} by F.~Bell where several interesting 
ideas are discussed. 
First of all, we would like to stress that the Unruh effect is 
not a phenomenon beyond QED, but rather a way of understanding
quantum effects via the non-trivial comparison of inertial and 
non-inertial frames.
Furthermore, the paper by F.~Bell is based on calculations 
(see, e.g., \cite{double}) where the polarizations 
are averaged over from the very beginning.
Therefore, it cannot address the polarization entanglement 
which played a crucial role in our analysis.
Indeed, expectation expressed at the end of {\tt arXiv:0809.1505v1},
namely that the polarizations of the two created photons are
always identical to the laser polarization 
(which in fact would mean no polarization entanglement),
is not correct.}. 
It should also be mentioned here that spontaneous parametric 
down-conversion (with x-rays \cite{x-ray} or in quantum optics) 
is based on the same diagrams as in Fig.~\ref{feynman3}.
However, in those cases the Unruh picture based on a nearly classical 
electron trajectory with a specific acceleration can no longer be 
applied and additional aspects (such as spatial phase matching) 
start to play a role.  

%%%%%%%%%%%%%%%%%%%%%%%%%%%%%%%%%%%%%%%%%%%%%%%%%%%%%%%%%%%%%%%%%%%%%%%%%%%%%%%
\section*{Acknowledgements}
%%%%%%%%%%%%%%%%%%%%%%%%%%%%%%%%%%%%%%%%%%%%%%%%%%%%%%%%%%%%%%%%%%%%%%%%%%%%%%%

R.~S.~acknowledges fruitful discussions with G.~Dunne, H.~Gies, D.~Habs, 
J.~Rafelski, G.~Schaller, V.~Serbo, P.~Thirolf, W.~G.~Unruh, 
and many others, in parts at the ELI Workshop and School on
{\em Fundamental Physics with Ultra-High Fields} 
Sep 28 - Oct 02 (2008) in Frauenw\"orth Monastery
(Frauenchiemsee, Germany). 
This work was supported by the DFG under grant SCHU~1557/1-2,3 
(Emmy-Noether program) and the Alexander von Humboldt Foundation. 
The authors  acknowledge the support by the European Commission under 
contract ELI pp 212105 in the framework of the program FP7 
Infrastructures-2007-1.

%%%%%%%%%%%%%%%%%%%%%%%%%%%%%%%%%%%%%%%%%%%%%%%%%%%%%%%%%%%%%%%%%%%%%%%%%%%%%%%

\end{document}